\documentclass{article}
\usepackage[utf8]{inputenc}
\usepackage{graphicx}
\usepackage{authblk}

\title{INDIGO-Datacloud: foundations and architectural description of a Platform as a Service oriented to scientific computing}

\author[1]{\small{D. Salomoni}}
\author[2]{I. Campos}
\author[3]{L. Gaido}
\author[4]{G. Donvito}
\author[4]{M. Antonacci}
\author[5]{P. Fuhrman}
\author[2]{J. Marco}
\author[2]{A. L\'opez-Garc\'{\i}a}
\author[2]{P. Orviz}
\author[6]{I. Blanquer}
\author[6]{M. Caballer}
\author[6]{G. Molt\'o}
\author[7]{M. Plociennik}
\author[7]{M. Owsiak}
\author[7]{M. Urbaniak}
\author[8]{M. Hardt}
\author[1]{A. Ceccanti}
\author[8]{B. Wegh}
\author[9]{J. Gomes}
\author[9]{M. David}
\author[1]{C. Aiftimiei}
\author[13]{L. Dutka}
\author[13]{B. Kryza}
\author[13]{T. Szepieniec}
\author[10]{S. Fiore}
\author[10]{G. Aloisio}
\author[11]{R. Barbera}
\author[11]{R. Bruno}
\author[11]{M. Fargetta}
\author[11]{E. Giorgio}
\author[12]{S. Reynaud}
\author[12]{L. Schwarz}
\author[14]{A. Dorigo}
\author[15]{T. Bell}
\author[15]{R. Rocha}

{\small {

\affil[1]{\small{INFN Division CNAF (Bologna), Italy}}
\affil[2]{IFCA, Consejo Superior de Investigaciones Cientificas-CSIC}
\affil[3]{INFN Division of Torino, Italy}
\affil[4]{INFN Division of Bari, Italy}
\affil[5]{Deutsches Elektronen Synchrotron (DESY), Germany}
\affil[6]{Universitat Polit\`ecnica de Val\`encia, Spain}
\affil[7]{PSNC IBCh PAS, Poland}
\affil[8]{Karlsruhe Institute of Technology (KIT), Germany}
\affil[9]{Laboratory of Instrumentation and Particles (LIP), Portugal}
\affil[10]{Fondazione Centro Euro-Mediterraneo sui Cambiamenti Climatici , Lecce, Italy}
\affil[11]{INFN Division of Catania, Italy}
\affil[12]{CC-IN2P3 / CNRS , France}
\affil[13]{Cyfronet AGH, Poland}
\affil[14]{INFN Division of Padova, Italy}
\affil[14]{CERN, IT Division, Geneva, Switzerland}
}}

\begin{document}

\maketitle

\begin{abstract}
In this paper we describe the architecture of a Platform as a Service (PaaS) oriented to computing and data analysis. In order to clarify the choices we made, we explain the features using practical examples, applied to several known usage patterns in the area of HEP computing. The proposed architecture is devised to provide researchers with a unified view of distributed computing infrastructures, focusing in facilitating seamless access. In this respect the Platform is able to profit from the most recent developments for computing and processing large amounts of data, and to exploit current storage and preservation technologies, with the appropriate mechanisms to ensure security and privacy.
\end{abstract}

\section{Introduction}

In the past decade, European research institutions, scientific collaborations and resource providers have been involved in the development of software frameworks that eventually led to the set-up of unprecedented distributed e-infrastructures such as the European Grid Infrastructure (EGI) \cite{EGI}; their collaboration made it possible to produce, store and analyze Petabytes of research data through hundreds of thousands of compute processors, in a way that has been instrumental for scientific research and discovery worldwide. 

This is particularly true in the area of High Energy Physics, where distributed computing has been instrumental as data analysis technology in the framework of the LHC computing Grid (WLCG) \cite{WLCG}.

New technological advancements, such as virtualization and cloud computing, pose new important challenges when it comes to exploit scientific computing resources. Services to researchers need to evolve in parallel to maximize effectiveness and efficiency, in order to satisfy different scenarios: from everyday necessities to new complex requirements coming from diverse scientific communities.  

In order to expose to researchers the power of current technological capabilities, a key challenge continues to be accessibility. Virtualization techniques have the potential to make available unprecedented amounts of resources to researchers. However, serving scientific computing is much more complex than providing access to a virtualized machine. It implies being able to support in a secure way complex simulations, data transfer \& analysis scenarios, which are by definition in constant evolution. 

This is particularly the case of the HEP areas, where we can distinguish three paradigmatic usage scenarios:

\begin{itemize}
\item Massive data analysis at the LHC and experiments which in general must deal with analysis of large amounts of data in computing farms. The paradigmatic case is the data analysis of the LHC at experiments like CMS and ATLAS. During LHC Run 1, the tools developed by WLCG oriented to Grid computing worked very smoothly. However, both the amount of data involved in Run 2 and beyond, and the evolution of the computing infrastructures implies that new scenarios for accessing resources need to be devised. The experiments CMS and ATLAS have both already analyzed the technical feasibility of using resources in cloud mode (see \cite{CMSCLOUD,ATLASCLOUD}). 

\item High Performance Computing in facilities with low latency interconnects dedicated to Monte Carlo Simulations, for example for Lattice Quantum Chromodynamics (Lattice QCD). Resource requirements for development phases are medium size clusters with up to a few hundred cores; in the production phase, Lattice QCD is served by large HPC farms which can require from a few thousand, up to tens of thousands of cores for ground breaking projects. The storage requirements are in the order of a few Terabytes for development and Petabyte for production phases \cite{LATTICEQCD}.
 
\item Phenomenological simulation \& prediction codes, often including legacy software components and complex library dependencies. The resources requirements can reach about a thousand cores, with storage in the range of the Terabyte. Here one current showstopper for researchers is the possibility of having installed the right environment (in terms of legacy libraries for example) \cite{PHENOTOOLS}. 
\end{itemize}

Making computing and storage resources really exploitable by scientists implies adapting access and usage procedures to the user needs. The challenge that remains is developing such procedures in a manner which is reliable, secure, sustainable, and with the guarantee that results are reproducible. 

In order to settle the scenario where our Platform is to be deployed, we must specify what we mean by the term resources. By this we always refer to a {\sl {large pool of computing and storage resources}}, in which the users do not need to know (because it is not relevant for the computation) which machine it is actually being used. If the computing requirements would have a strong hardware dependency (optimized for a particular processor), or if the application is so large scale that a significant fraction of the available resources in a given computing center needs to be used, the discussion would be completely different. 

This is the context of the work of the HORIZON 2020 project INDIGO-DataCloud, referred to as INDIGO from now on \cite{INDIGO}, addressing the challenge of developing advanced software layers, deployable in the form of a data/computing platform, targeted at scientific communities.  

In this paper we present an architectural design to satisfy the challenges mentioned above, and which we have developed in the framework of the INDIGO project. We highlight the interrelations among the different components involved and outline the reasoning behind the choices that were made. 

The remainder of this paper is structured as follows. 

We first summarize the proposed progress beyond the state of the art (section \ref{sec:summary}). In order to motivate our choices we next describe several generic user scenarios in Section \ref{sec:motivation}, with the main functionalities we want to satisfy, offering a global, high-level view of the INDIGO architecture highlighting the interrelations among the different layers. Section \ref{sec:paas} describes the architecture of the PaaS layer, together with the description of the user interfaces to be developed. Section \ref{sec:infrastructure} is devoted to the computer center layer of the architecture. In particular, it describes the middleware choices and developments needed to fully support the PaaS layer at the infrastructure level. Section \ref{sec:portals} describes how INDIGO interfaces with users. Finally, section \ref{sec:data} describes the solutions for unified data management.
 
The paper is concluded by Section \ref{sec:conclusions}, drawing some conclusions and highlighting future work.

\section{Proposed progress beyond the state-of-the-art}
\label{sec:summary}

In order to provide the reader with a global understanding of the practical implications of this work, we highlight here the technical progress we intend to achieve at the different layers that compose a scientific computing infrastructure.

\subsection{Infrastructure layer}

At the resource provider level, computing centers offer resources in an Infrastructure as a Service (IaaS) mode. The main challenges to be addressed at this level are:

\begin{enumerate}
\item Improved scheduling for allocation of resources by popular open source Cloud platforms, i.e. OpenStack \cite{OPENSTACK} and OpenNebula \cite{OPENNEBULA}.

In particular, both better scheduling algorithms and support for spot-instances are currently much needed. The latter are in particular needed to support allocation mechanisms similar to those available on commercial clouds such as Amazon Web Services and Google Cloud Platform, while the former are useful to address the fact that typically all computing resources in scientific data centers are always in use. 

For LHC data analysis several attempts at using such spot-instances in the framework of commercial cloud providers have already taken place in the experiment CMS \cite{CMSCLOUD} and ATLAS \cite{ATLASCLOUD}. Having an open source software framework supporting such operation mode at the IaaS level will therefore be very important for the centers aiming to support such data analysis during LHC Run 2 and beyond.

\item Improved Quality of Service (QoS) capabilities of storage resources.  The challenge here is to develop a better support for quality of services in storage, to enable high-level storage management systems (such as FTS) and make it aware of information about the underlying storage qualities.

The impact of such QoS when applied to storage interfaces such as dCache will be obvious for LHC Data analysis and for the long-term support, preservation and access of experiment data.

For example can we save a lot of redundant copies, when the high level storage manager knows how many copies are already available at one storage location. Due to a lack of this information the assumption made is that only one copy is available.

\item Improved capabilities for networking support. This is particularly the case when it comes to deploy tailored network configurations in the framework of OpenNebula and OpenStack.

\item Improved and transparent support for Docker containers. 

Containers provide an easy and efficient way to encapsulate and transport applications. Indeed, they represent a higher level of abstraction than the crude concept of a "virtual machine". The benefits of using containers, in terms of easiness in the deployment of specialized software, including contextualization features, eg. for phenomenology applications, are clear.

They offer obvious advantages in terms of performance when compared with virtual machines, while also opening the door to exploit specialized hardware such as GPGPUs and low-latency interconnection interfaces (InfiniBand).

The general idea here is to make containers "first-class citizens" in scientific computing infrastructures. 

\end{enumerate}

\subsection{Platform layer}

In the next layer we find the PaaS layer. This is a set of services whose objective is to leverage disparate hardware resources coming from the IaaS level (Grid of distributed clusters, public and private clouds, HPC systems) to enhance the user experience. 

In this context a PaaS should provide advanced tools for computing and for processing large amounts of data, and to exploit current storage and preservation technologies, with the appropriate mechanisms to ensure security and privacy. The following points describe the most important missing capabilities which today require further developments:

\begin{enumerate}

\item Improved capabilities in the geographical exploitation of Cloud resources.
End users do not need to know where resources are located, because the PaaS layer should be hiding the complexity of both scheduling and brokering. 

\item Support for data requirements in Cloud resource allocations.
Resources can be allocated where data is stored, therefore facilitating interactive processing of data. The benefits of such an enhancement are clear for software stacks for interactive data processing tools such as ROOT \cite{ROOT} and PROOF \cite{PROOF}.

\item Support for application requirements in Cloud resource allocations. For example, a given user can request to deploy an application on a cluster with Infiniband interfaces, or with access to specialized hardware such as GPGPUs. Elasticity in the provisioning of such specialized small size clusters for development purposes would have a great impact in the everyday work of many researchers in the area of Lattice QCD for example.

\item Transparent client-side import/export of distributed Cloud data.

\item Deployment, monitoring and automatic scalability of existing applications, including batch systems on-demand.
For example, existing applications such as web front-ends, PROOF clusters or even a complete batch system cluster (with appropriate user interfaces) can be automatically and dynamically deployed in highly-available and scalable configurations.

\item Integrated support for high-performance Big Data analytics and workflow engines such as Taverna \cite{TAVERNA}, Ophidia \cite{OPHIDIA} or Spark \cite{SPARK}.

\item Support for dynamic and elastic clusters of computational resources.

\end{enumerate}

\subsection{User interface layer}

In the next layer we find the user interface, which is responsible to convey all the above mentioned developments to the user. This means in particular that it should provide ready-to-use tools for such capabilities to be exploited, with the smoothest possible learning curve. 

Providing such an interface between the user and the infrastructure poses two fundamental challenges:

\begin{enumerate}
    
\item Enabling infrastructure services to accept state of the art user authentication mechanisms (e.g. OpenID connect, SAML) on top of the already existing X.509 technology. For example, distributed authorization policies are very much needed in scientific cloud computing environments, therefore a dedicated development effort is needed in this area. Hence, the Authentication and Authorization Infrastructure (AAI) is a key ingredient to be fed into the architecture.

\item Making available the appropriate libraries, servlets and portlets, implementing the different functionalities of the platform (AAI, data access, job processing, etc.) that are the basis to integrate such services with known user tools, portals and mobile applications.

\end{enumerate}

\section{Motivation and High Level view of the Architecture}
\label{sec:motivation}

We have designed an architecture containing the elements needed to provide scientific users with the capability of using heterogeneous infrastructures, adressing the challenges described above. In the following we describe the rational and motivations of the technical choices we made.

As a first step we have performed a detailed user requirements analysis, whose main conclusions we show in the form of two generic scenarios: the first is computing oriented, while the second is data analysis oriented. For full details containing user communities description and detailed usage patterns we refer to our requirements document \cite{D21}.

\subsection{Pilot user scenarios}

\begin{figure}
  \centering
  \includegraphics[width=11cm]{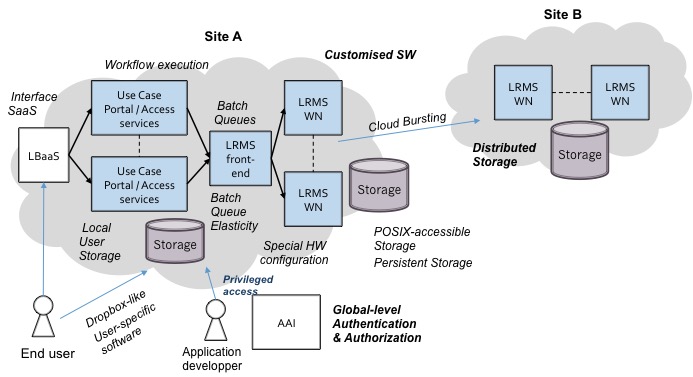}
  \caption{User Community Portal Service}
  \label{fig:1}
\end{figure}

Our architecture is based on the analysis of a number of use cases originating from different research communities in the areas of High Energy Physics, Environmental modelling, Bioinformatics, Astrophysics, Social sciences and others. From this requirements analysis we have extracted two generic usage scenarios, which can support a wide range of applications in these areas.

The first generic user scenario is a computing portal service. In such scenario, computing applications are stored by the application developers in repositories as downloadable images (in the form of VMs or containers). Such images can be accessed by users via a portal, and require a back-end for execution; in the most common situation this is typically a batch queue.  

The number of nodes available for computing should increase (scale out) and decrease (scale in), according to the workload. The system should also be able to do Cloud-bursting to external infrastructures when the workload demands it. Furthermore, users should be able to access and reference data, and also to provide their local data for the runs. A solution along these lines is shown in Figure \ref{fig:1}.

\begin{figure}
  \centering
  \includegraphics[width=11cm]{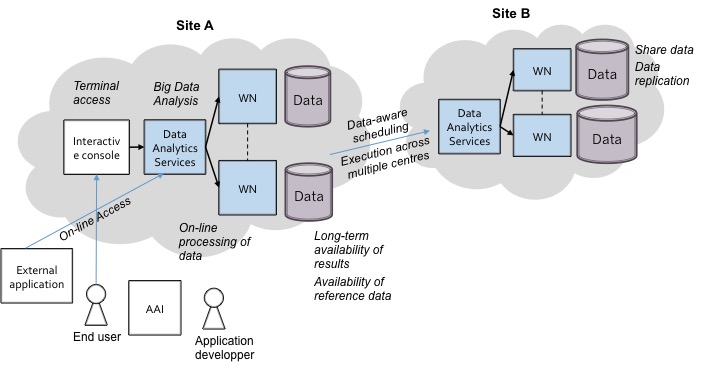}
  \caption{Data Analysis Service}
  \label{fig:2}
\end{figure}

A second generic use case is described by scientific communities that have a coordinated set of data repositories and software services (for example PROOF, or R-studio) to access, process and inspect them. Processing is typically interactive, requiring access to a console deployed on the data premises. In Figure \ref{fig:2} we show a schematic view of such a use case.

As pointed out in the introduction, the current technology based on lightweight containers and related virtualization developments make it possible to design software layers in the form of platforms that support such usage scenarios in a relatively straightforward way. 

We can see already many examples in the industrial sector, in which open source PaaS solutions such as OpenShift or Cloud Foundry are being deployed to support enterprise work in different sectors \cite{REDHAT}. 

However, the case of supporting scientific users is more complex, first because of the heterogeneous nature of the infrastructures at the IaaS level (i.e. the resource centers), and secondly because of the inherent complexity of the scientific work requirements. The key point here is to find the right agreement to unify interfaces between the PaaS and IaaS levels.

\subsection{PaaS layers over production scientific e-Infrastructures}

For the architecture to go beyond just a theoretical implementation of tools and APIs, we must include the practicalities of the computing centers in the discussion. The architecture should be capable of supporting the interaction with the resource centers via standard interfaces. Here the word standard is meant in a very wide sense including \textit{de jure} as well as \textit{de facto} standards. 

Virtualization of resources is the key word in order to properly address the interface with the resource centers. In other words, the software stack should be able to virtualize local compute, storage and networking IaaS resources, providing those resources in a standardized, reliable and performing way to remote customers or to higher level federated services. 

The IaaS layer is normally provided to scientists by large resource centers, typically engaged in well-established European e-infrastructures. The e-infrastructure management bodies or the resource centers themselves will select the components they operate. Therefore, the success of any software layer in this respect is being able to be flexible enough as to interact with the most popular choices of the computer centers, without interfering, or very minimally, in the operation of their facilities.

\begin{figure}
  \centering
  \includegraphics[width=10cm]{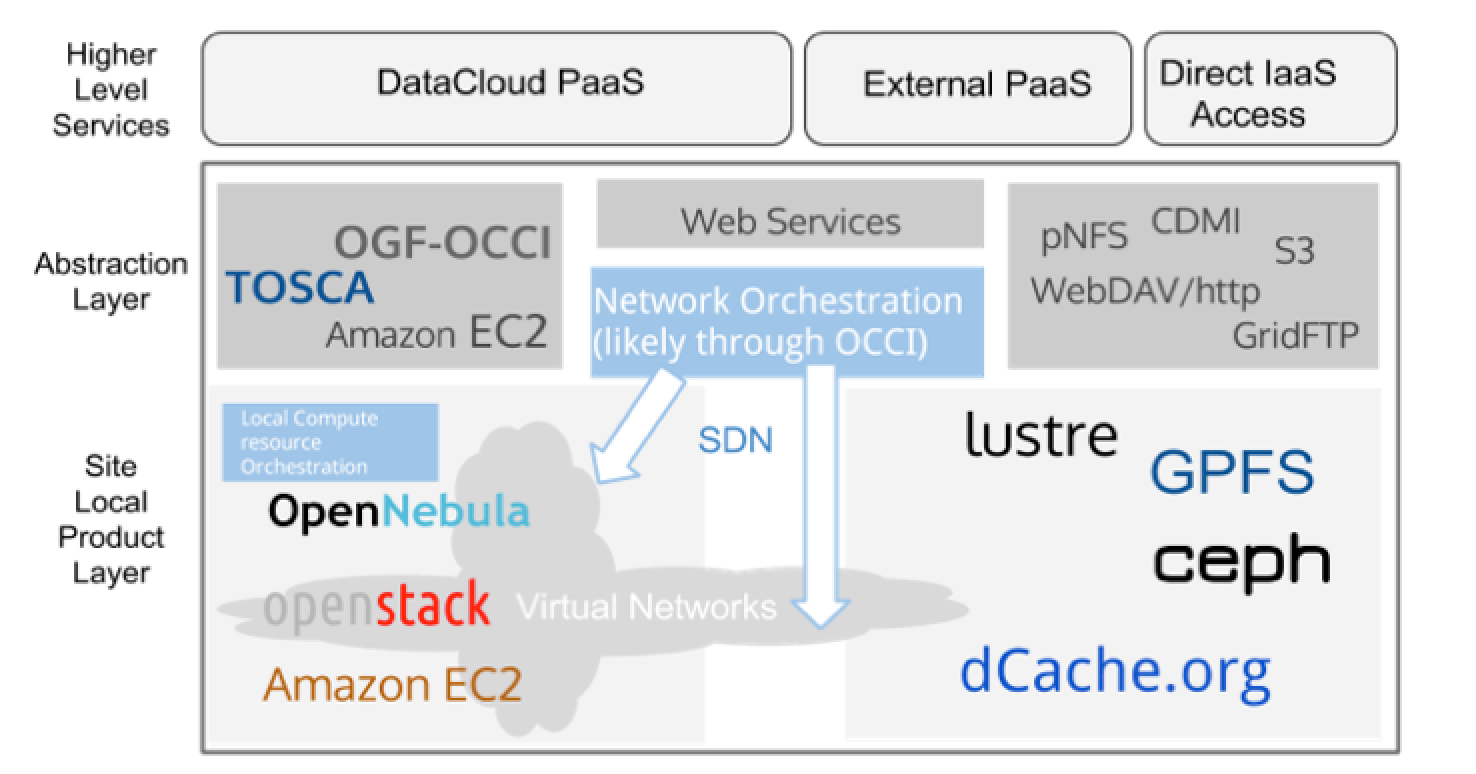}
  \caption{High-level view of the interaction between the IaaS and the PaaS layer}
  \label{fig:3}
\end{figure}

As a consequence, as a part of the development effort, we have analyzed a selection of the most prominent components to interface computing and storage in the resource centers, and develop the appropriate interfaces to high-level services based on standards. Figure \ref{fig:3} shows a schematic view of the interrelation among those components.

The PaaS core components will be deployed as a suite of small services using the concept of “micro-service” . This term refers to a software architecture style, in which complex applications are composed of small independent processes communicating with each other via lightweight mechanisms like HTTP resource APIs. The modularity of micro-services makes the approach highly desirable for architectural design of complex systems, where many developers are involved.

\begin{figure}
  \centering
  \includegraphics[width=12cm]{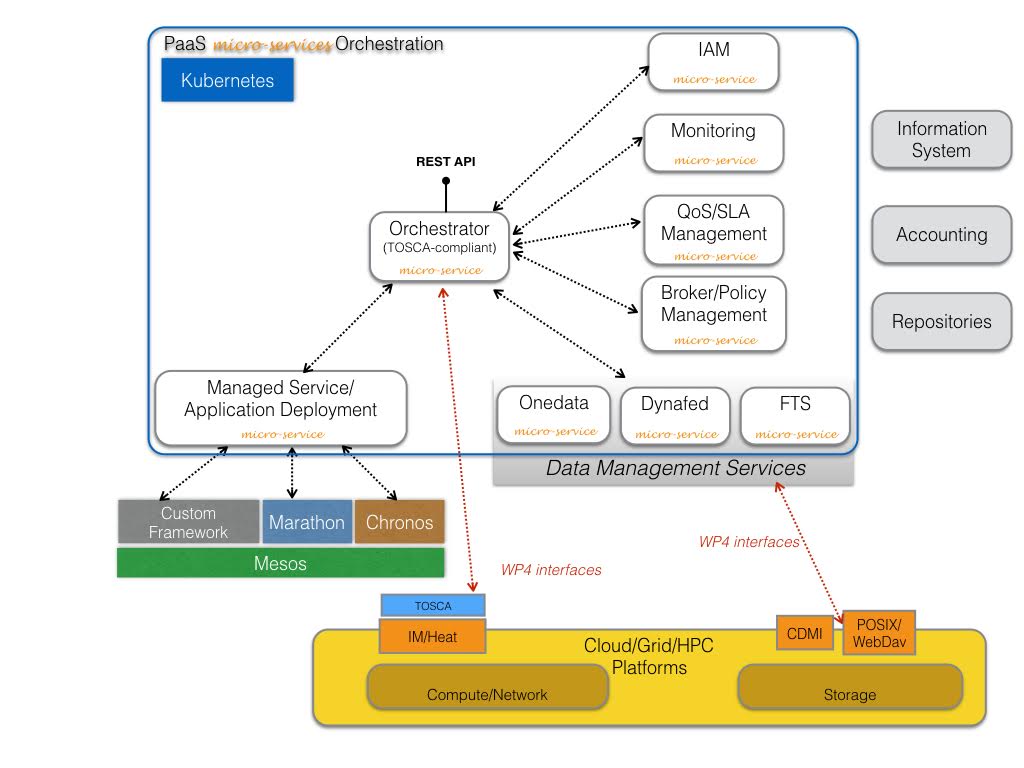}
  \caption{Key components of the PaaS and their high-level interrelations}
  \label{fig:4}
\end{figure}

Kubernetes \cite{KUBERNETES}, an open source platform to orchestrate and manage Docker \cite{DOCKER} containers, will be used to coordinate the micro-services in the PaaS. Kubernetes is extremely useful for the monitoring and scaling of the services, and will ensure the reliability of all of them. In Figure \ref{fig:4} we show the high-level view of the PaaS in which the interrelations among services are also indicated with arrows.

The following list briefly describes the key components of the INDIGO PaaS:

\begin{itemize}
    
\item  the Orchestrator: this is the core component of the PaaS layer. It receives high-level deployment requests from the user interface software layer, and coordinates the deployment process over the IaaS platforms;

\item the Identity and Access Management (IAM) Service: it provides a layer where identities, enrolment, group membership, attributes and policies to access distributed resources and services can be managed in a homogeneous and interoperable way;   

\item the Monitoring Service: this component is in charge of collecting monitoring data from the targeted clouds, analysing and transforming them into information to be consumed by the Orchestrator; 

\item the Brokering/Policy Service: this is a rule-based engine that allows to manage the ranking among the resources that are available to fulfil the requested services. The Orchestrator will provide the list of IaaS instances and their properties to the Rule Engine. The Rule Engine will then be able to use these properties in order to choose the best site that could support the users’ requirements. The Rule Engine can be configured with different rules in order to customize the ranking;    

\item the QoS/SLA Management Service: it allows the handshake between a user and a site on a given SLA; moreover, it describes the QoS that a specific user/group has, both over a given site or generally in the PaaS as a whole. This includes a priority for a given user, i.e. the capability to access different levels of QoS at each site (e.g., Gold, Silver, Bronze services);

\item the QoS/SLA Management Service: it allows the handshake between a user and a site on a given SLA; moreover, it describes the QoS that a specific user/group has, both over a given site or generally in the PaaS as a whole. This includes information about the actual service quality of storage spaces and stored files at endpoints plus the possibility to change these service qualities for stored data.

\item the Managed Service/Application (MSA) Deployment Service: it is in charge of scheduling, spawning, executing and monitoring applications and services on a distributed infrastructure; it is implemented as a workflow programmatically created and executed by the Orchestrator, as detailed in the next section. 

\item the Infrastructure Manager (IM) \cite{IM}: it deploys complex and customized virtual infrastructures on IaaS Cloud deployment providing an abstraction layer to define and provision resources in different clouds and virtualization platforms;

\item the Data Management Services: this is a collection of services that provide an abstraction layer for accessing data storage in a unified and federated way. These services will also provide the capabilities of importing data, schedule transfers of data, provide a unified view on QoS and distributed Data Life Cycle Management. 

\end{itemize}

Figure \ref{fig:4} shows also the interaction between the MSA core service and the external components Apache Mesos, Marathon and Chronos \cite{MESOS,MARATHON,CHRONOS}. These are open-source components that have been selected after a deep analysis of the cutting edge technologies for application and container management.

Mesos is a smart resource manager originally conceived as research project at UC Berkeley and currently used in production by the industrial sector as well.

Mesos abstracts CPU, memory, storage and other compute resources away from machines (physical or virtual) and allows sharing them across different distributed applications (called frameworks). Sophisticated two-level scheduling and efficient resource isolation are the key-features of this middleware that are exploited in the INDIGO PaaS, in order to run different workloads (long-running services, batch jobs, etc) on the same resources while preserving isolation and prioritizing their execution. 

The Mesos cluster architecture \cite{MESOS-IMAGE} is organized in two sets of nodes: masters, which coordinate the work, and slaves, which execute it. The master nodes are responsible for handling the resources available on the slaves and offer them to the frameworks according to specific policies; then the frameworks are responsible for the application specific scheduling policy. This allows for more fine-tuned scheduling and dynamic partitioning of resources and for application aware scheduling.

The MSA service is implemented as a complex workflow managed by the Orchestrator that delegates to two already available Mesos frameworks for deploying containers on the IaaS sites: Marathon, which allows to deploy and manage Long-Running Services, and Chronos, which allows to execute jobs. 

The capabilities of Mesos and its frameworks will be enhanced by adding crucial features like: the elasticity of the Mesos cluster that will automatically shrink or expand depending on the tasks queue, as detailed in the next section, the automatic scaling of the user services that run on top of the Mesos cluster, a stronger authentication mechanism based on OpenID Connect.

\section{Architecture of the PaaS Layer}
\label{sec:paas}

Generally speaking, a Platform as a Service (PaaS) is a software suite, which is able to receive programmatic resource requests from end users, and execute these requests provisioning the resources on some e-infrastructures.

In the INDIGO approach, the PaaS will deal with the instantiation of services and with application execution upon user requests relying on the concept of micro-services. In turn, the micro-services will be managed using Kubernetes, in order, for example, to select the right end-point for the deployment of applications or services. Cross-site deployments will also be possible.  

The language in which the PaaS is going to receive end user requests is TOSCA \cite{TOSCA} (Topology and Orchestration Specification for Cloud Applications). It is an OASIS specification for the interoperable description of application and infrastructure cloud services, the relationships between parts of these services, and their operational behaviour. In particular we will be using the TOSCA simple profile in YAML Version 1.0 \footnote{http://docs.oasis-open.org/tosca/TOSCA-Simple-Profile-YAML/v1.0/csprd01/TOSCA-Simple-Profile-YAML-v1.0-csprd01.html}.

TOSCA has been selected as the language for describing applications, due to the wide-ranging adoption of this standard, and since it can be used as the orchestration language for both OpenNebula (through the IM \cite{IM}) and OpenStack (through Heat \cite{HEAT}).

The PaaS Core provides an entry point to its functionality via the Orchestrator service, which features a RESTful API that receives a TOSCA-compliant description of the application architecture to be deployed.  Providing such TOSCA-compliance enhances interoperability with existing and prospective software. 

Users can choose between accessing the PaaS core directly or using a Graphical User Interface or simple APIs. A user authenticated on the INDIGO Platform will be able to access and customize a rich set of TOSCA-compliant templates through a GUI-based portlet.

The INDIGO repository will provide a catalogue of pre-configured TOSCA templates to be used for the deployment of a wide range of applications and services, customizable with different requirements of scalability, reliability and performance. 

\begin{figure}
  \centering
  \includegraphics[width=12cm]{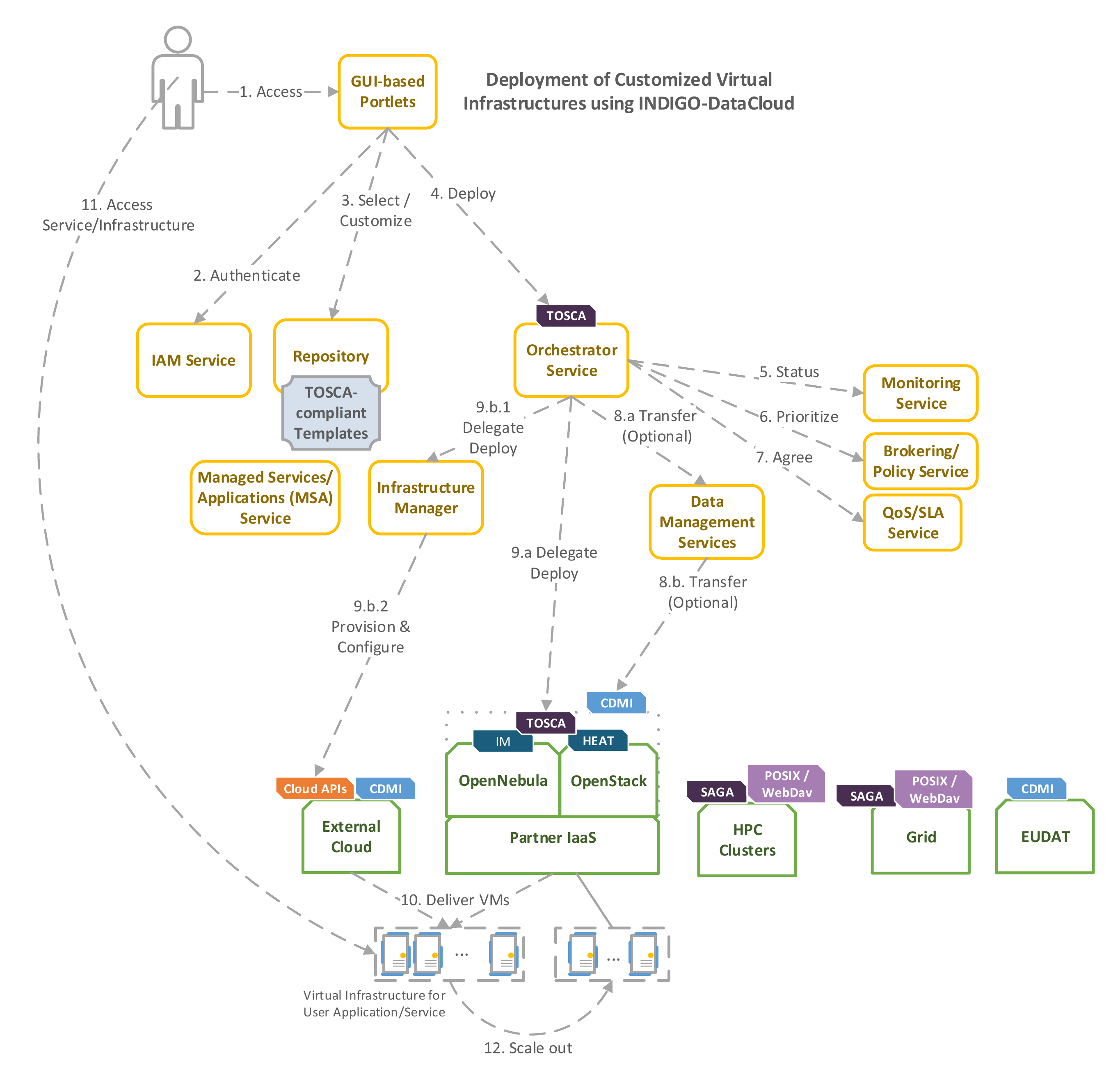}
  \caption{Deployment of a customized virtual infrastructure: When a customized virtual infrastructure deployment is requested (scenario A), the Orchestrator manages the instantiation and configuration of the required resources (e.g. virtual machines) on the selected IaaS infrastructure using the REST APIs exposed by the IaaS orchestrator (i.e. Heat or IM) of the INDIGO sites or delegating the interaction with external clouds to a dedicated instance of the IM.}
  \label{fig:5}
\end{figure}

\begin{figure}
  \centering
  \includegraphics[width=12cm]{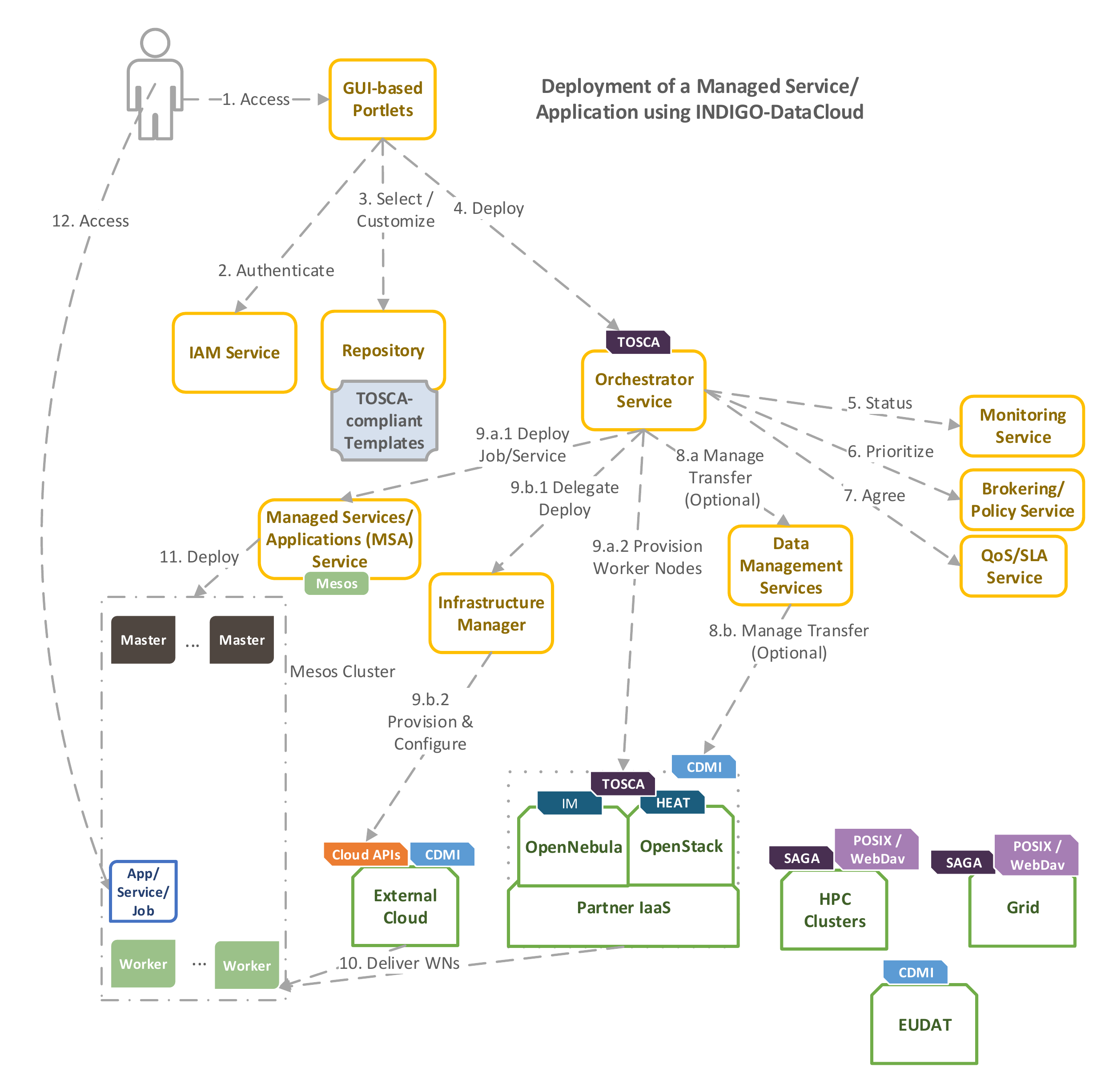}
  \caption{Deployment of a managed service/application: When a managed PaaS service deployment is requested (scenario B), the Orchestrator interacts with the Managed Service/Application (MSA) Deployment Service in order to supervise its deployment on the elastic Mesos cluster that will host the user application/service. }
  \label{fig:6}
\end{figure}

In these templates a user can choose between two different examples of generic scenarios:

\begin{itemize}
    \item Scenario A. Deploy a customized virtual infrastructure starting from a TOSCA template that has been imported, or built from scratch (see Figure \ref{fig:5}). The user will be able to access the deployed customized virtual infrastructure and run/administer/manage applications running on it.
    \item Scenario B. Deploy a service/application whose life-cycle will be directly managed by the PaaS platform (see Figure \ref{fig:6}). The user will be returned the list of endpoints to access the deployed services.
\end{itemize}

In both cases the selected template can be submitted to the PaaS Orchestrator using its REST API endpoint. Then, the Orchestrator collects all the information needed to generate the deployment workflow:

\begin{itemize}
\item Health status and capabilities of the underlying IaaS platforms and their resource availability from the Monitoring Service;
\item Priority list of sites sorted by the Brokering/Policy Service on the basis of rules defined per user/group/use-case;
\item QoS/SLA constraints from the SLA Management System;
\item The status of the data files and storage resources needed by the service/application and managed by the Data Management Service.
\end{itemize}

This information is used to perform the matchmaking process and to decide where to deploy each service. Note that the Orchestrator is able to trigger the data migration function provided by the Data Management Service component if the data location does not meet the application deployment requirements.

As pointed out before, the MSA is implemented as a deployment plan managed by the workflow engine provided by the Orchestrator, that, starting from the TOSCA template that describes the deployment request, creates the workflow programmatically.

The MSA workflow relies on the capabilities of Mesos to manage the distributed set of IaaS resources: its execution is accomplished through a set of calls to the APIs endpoints of the Mesos cluster, whose architecture consists of one or more master nodes, and of slave nodes that register with the master and offer resources from the IaaS nodes.

The master node is aware of the state of the whole IaaS resources, and can share and assign them to the different applications (called frameworks in the Mesos terminology) according to specific scheduling policies.  Mesos provides a default scheduling algorithm, the Dominant Resource Fairness (DRF)\cite{DRF}, but it is possible to develop custom algorithms and easily configure Mesos to use them thanks to its modular (plugin) architecture.

The Automatic Scaling Service, based on EC3/CLUES \cite{CLUES}, ensures the elasticity and scalability of the Mesos cluster by monitoring its status. When additional computing resources (worker nodes) are needed, the Orchestrator will be requested to deploy them on the underlying IaaS matching the QoS/SLA, health and user/group/use-case policies selected by the Broker. 

In the case of Long-Running Services, the Management Service/Application (MSA) deployment Service will use Marathon \cite{MARATHON} (a container orchestration platform available in the Mesos framework) to ensure that the services are always up and running. Marathon is able to restart the services, migrate them if problems occur, handle their mutual dependencies and load-balancing, etc.

The MSA Deployment Service will also use Chronos \cite{CHRONOS} (a fault tolerant scheduler available in the Mesos framework) to execute applications having input/output requirements or dependencies. It may also handle the rescheduling of failed applications, or simple workflows composed by different applications.

By leveraging the Mesos plugin-based architecture, new frameworks can be developed, such as the one able to deploy a batch cluster (e.g. HTCondor) on demand, in order to meet specific use-cases. For example, batch execution of LHC data analysis is often using HTCondor to manage the job scheduling \cite{LHCCondor}

With respect to Data Management Services, some interfaces are provided to advanced users for specific data management tasks.

First of all, the OneData \cite{ONEDATA} component provides several features: a web-based interface for managing user spaces (virtual folders) and controlling access rights to files on a fine-grained level, a Posix interface to a unified file system namespace addressing both local and remote data, caching capabilities and the possibility to map object stores such as S3 \cite{S3} to Posix filesystems. Additionally, the FTS-3 \cite{FTS} service will provide a web-based interface for monitoring and scheduling large data transfers. 

Furthermore, all the standard interfaces exposed by the data management components will be accessible to users’ applications as well through standard protocols such as CDMI \cite{CDMI} and WebDAV \cite{WEBDAV}.

\section{Architectural impact on the Infrastructure}
\label{sec:infrastructure}

The impact of the implementation of INDIGO software developments at the level of infrastructure resource providers is a key discussion to guarantee the adoption of the solutions being developed.

A successful architecture should be able to provide the means to unify the interfaces between the PaaS layer and the core services. This is necessary, as resources sites have already their own administration software installed. A PaaS, like any software layer dealing with resource management, needs to be totally customizable to guarantee a good level of adoption by infrastructure providers.

Therefore our strategy is to focus on the most popular standards and provide well-documented common interfaces to these. Examples are OCCI \cite{OCCI}, TOSCA or a consistent container support for OpenStack and OpenNebula. An example in the data area is the support of CDMI for the various storage systems like dCache, StoRM, HPSS and GPFS. 

A closely related goal, often being the result of the topic discussed previously, is the functional unification between different software systems.
Support is being introduced in the IM to be able to deploy application architectures described in TOSCA on the different Cloud back-ends supported by the IM, including OpenNebula, OpenStack and public Cloud providers. In particular, for OpenStack,  the heat-translator project will be employed to deploy TOSCA-compliant architectural descriptions on OpenStack sites

Another area of development which is currently demanded by scientific communities is the introduction 
of "Quality of Service" and "Data Life-cycle Policies" in the data area. This is the result of the various 
"Data Management Plans, DMP" provided by data intensive communities and also required by the European Commision (EC) when submitting proposals. One important aspect of DMPs is the handling of quality of service and access control of 
precious and irreproducible data over time, resulting in support and manageability of those attributes at 
the site or storage system level.

Although the different types of resources are closely interlinked, we distinguish between Computing, Storage and Network resources for organizational reasons.

In the computing area the provision of standard APIs is covered by supporting OCCI at the lowest resource management level and TOSCA at the infrastructure orchestration level. 

Within the storage area, common access and control mechanisms are evaluated for negotiating data quality properties, e.g. access latency and retention policies, as well as the orchestration of data life cycles for archival. Together with established standardization bodies, like RDA\cite{RDA} and OGF\cite{OGF}, we envision to extend the SNIA CDMI protocol\cite{CDMI} for our purposes. 

Similarly for Networking, we need to evaluate commonalities in the use of Software Defined Networks (SDN) between different vendors of network appliances.

One notably attractive concept is that all features developed at the IaaS level will not only be available through the INDIGO PaaS layer, but can be utilized by users accessing the IaaS layer directly. 

Similarly, tracking of user identities is available throughout the entire execution stack. Consequently, users can be monitored down to the IaaS layer with the original identities they provided to portals or workflow engines when logged via the PaaS INDIGO layer.

\subsection{Software to interface the Resource Centers with the PaaS}

Based on the scientific use cases we have considered (see \cite{D21}), we identified a set of features that have the potential to impact in a positive way the usability and easy access to the Infrastructure layers. 

\begin{figure}
  \centering
  \includegraphics[width=10cm]{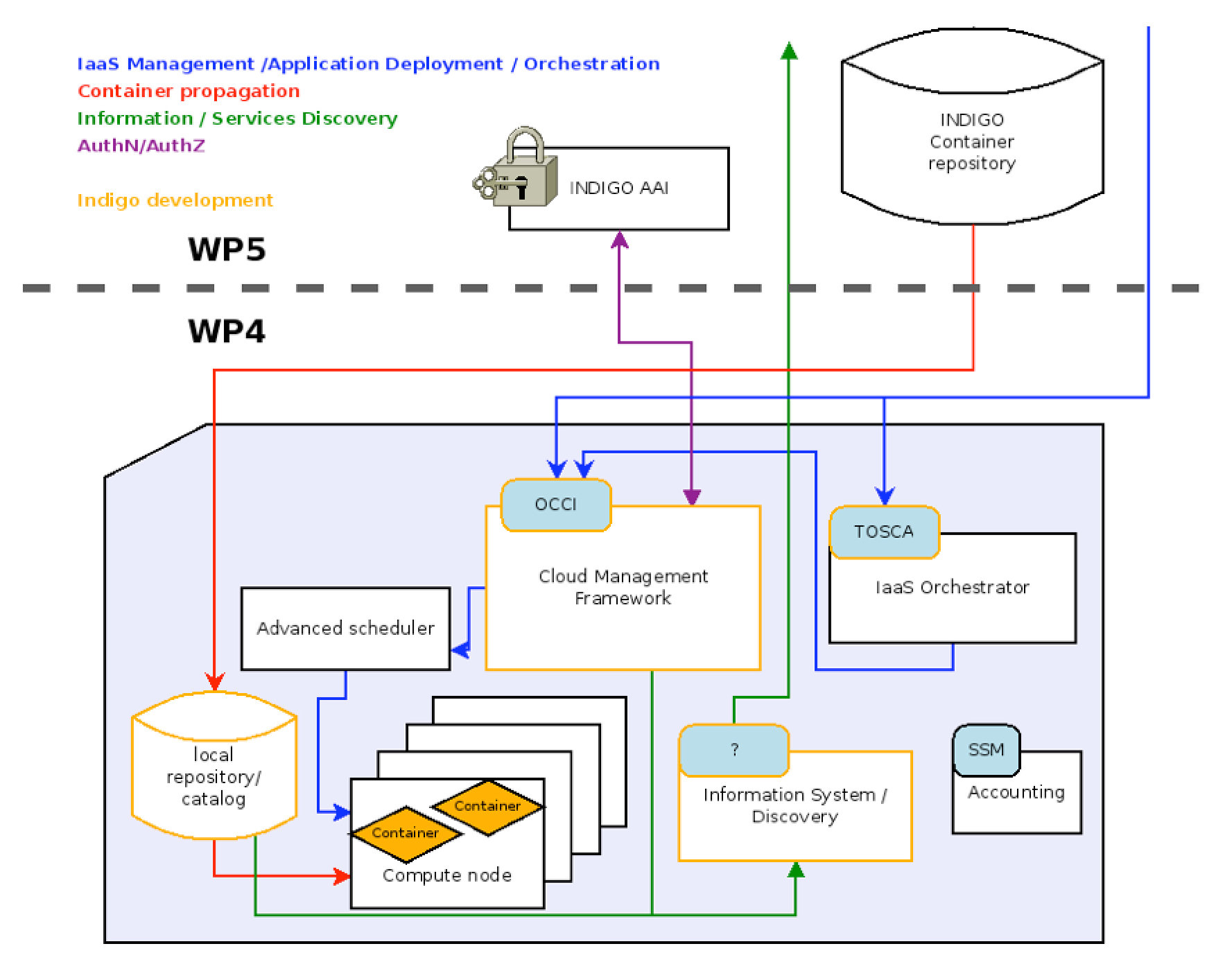}
  \caption{Infrastructure view of the INDIGO architecture}
  \label{fig:8}
\end{figure}

In the computing area, these features are enhanced support for containers, integration of batch systems, including access to hardware specific features like InfiniBand and General Purpose GPUs, support for trusted container repositories, introduction of spot instances and fair-share scheduling for selected Cloud Management Frameworks (CMF), as well as orchestration capabilities common to INDIGO selected CMFs using TOSCA. See Figure \ref{fig:8} for a graphical representation.

In certain applications, the use of ‘Containers’ as a lightweight alternative to hypervisor-based virtualization is becoming extremely popular, due to their significantly lower overhead. However, support in major Cloud Management Frameworks (CMFs) is still under development or does not exist at all. 

For OpenStack and OpenNebula, the top two CMF’s on the market, INDIGO, in collaboration with the corresponding Open Source communities, is spending significant efforts to make containers first-class citizens and, concerning APIs and management, indistinguishable from traditional VMs.  

While in OpenStack integration of Nova-Docker will introduce support for Docker containers, for OpenNebula, additional developments are required. In particular,we have developed OneDock \cite{ONEDOCK}, which introduces Docker as an additional hypervisor for OpenNebula, maintaining full integration with the OpenNebula APIs and web-based portal (SunStone)

Although cloud-like access to resources is becoming popular and cloud middleware is being widely deployed, traditional scientific data centers still provide their computational power by means of Batch Systems for HTC and HPC. Consequently, it is interesting to facilitate the integration of containers in batch systems, providing users with the ability to execute large workloads embedded inside a container. 

With the pressure of optimizing computer center resources but at the same time providing fair, traceable and legally reproducible services to customers, available cloud schedulers need to be improved. Therefore, we are focusing on the support of spot-instances allowing brokering resources based on SLAs and prices. Technically this feature requires the CMF to be able to preempt active instances based on priorities. 

On the other hand, to guarantee an agreed usage of compute cycles integrated over a time interval, we need to invest in the evaluation and development of fair-share schedulers integrated in CMFs. This requires a precise recording of already used cycles and the corresponding readjustment of permitted current and future usage per individual or group. The combination of both features allows resource providers to partition their resources in a dynamic way, ensuring an optimized utilization of their infrastructures.
					
The middleware also provides local site orchestration features by adopting the TOSCA standard in both OpenStack and OpenNebula, with similar and comparable functionalities.

Finally, resource orchestration is also covered within this architecture. Although this area can be managed at a higher level, we will provide compute, network and storage resource orchestration by means of the TOSCA language standard at the IaaS level as well. As a result, both, the upper platform layer and the infrastructure user may deploy and manage complex configurations of resources more easily. 

While in the cloud computing area, the specification of service qualities, e.g. number and power of CPUs, the amount of RAM and the performance of network interfaces, is already common sense, negotiating fine grained quality of service in the storage area, in a uniquely defined way, is not offered yet. 

\begin{figure}
  \centering
  \includegraphics[width=10cm]{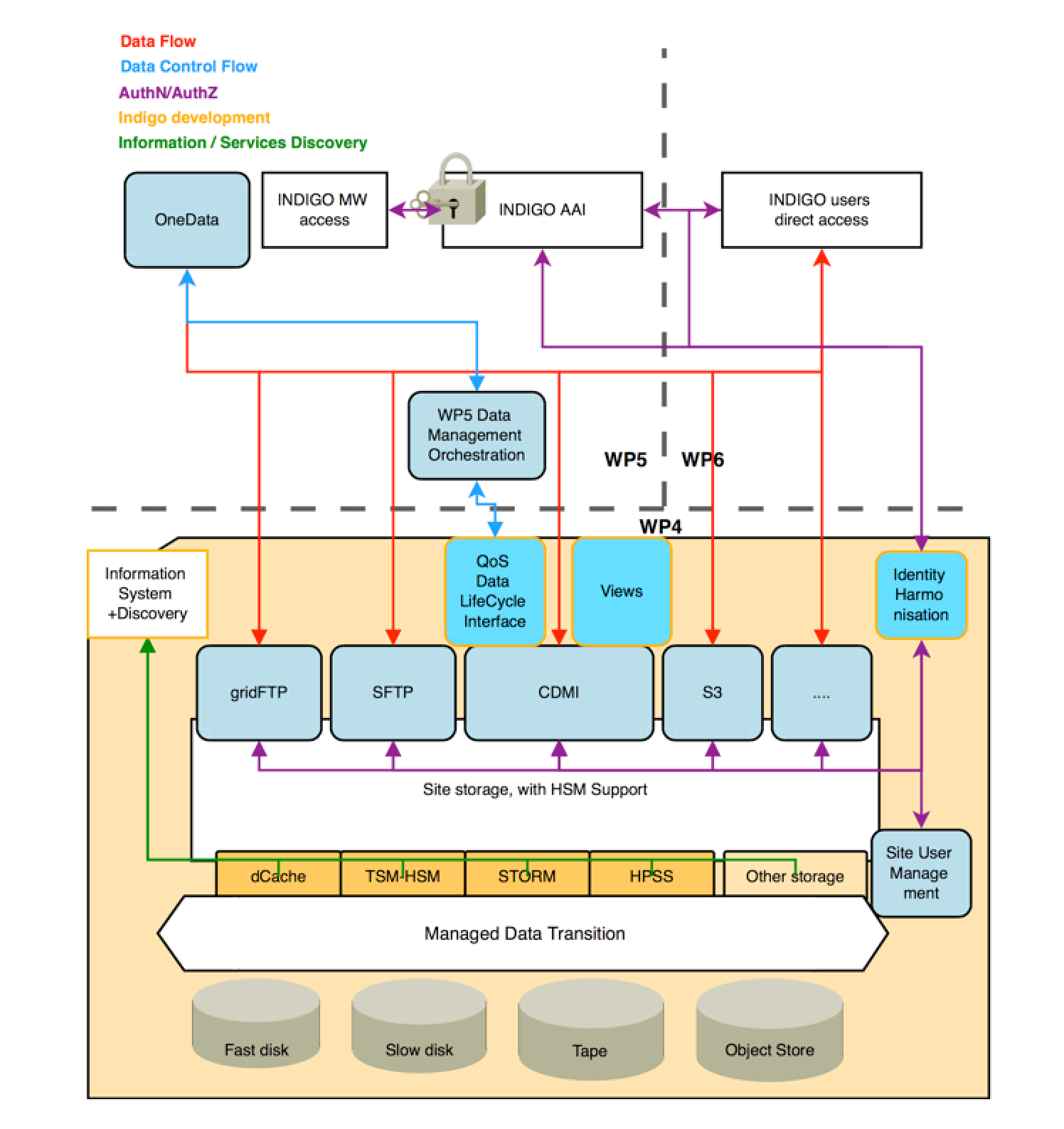}
  \caption{Storage services from the INDIGO architecture perspective}
  \label{fig:9}
\end{figure}

Therefore, the high level objective of the storage area is to establish a standardized interface for the management of Quality of Services (QoS) and Data Life Cycle in Storage (DLC). Users of e-infrastructures will be enabled to query and control properties of storage areas, like access latency, retention policy and migration policies with one standardized interface. A graphical representation of the components is shown in Figure \ref{fig:9}.

To engage scientific communities into this endeavour as early as possible, INDIGO initiated a working group within the framework of the Data Research Alliance \cite{RDA} (RDA), and will incorporate ideas and suggestions of that group at any stage of the project into the development of the system\footnote{See https://owncloud.indigo-datacloud.eu/index.php/s/Ur7ORPKM8lF0adQ}.

As with all infrastructure services, the interface is supposed to be used by either the PaaS storage federation layer or by user applications utilizing the infrastructure directly.

This will be pursued in a component-wise approach. Development will focus on QoS and interfaces for existing storage components and transfer protocols that are available at the computer centers. Ideally,  the Storage QoS component can be integrated just like another additional component into existing infrastructures.
					
Besides providing the translation layer between the management interface and the underlying storage technologies, the software stack needs to be integrated into an existing local infrastructure.

The high level objective of the network area is to provide mechanisms for orchestrating local and federated network topologies. To unify the orchestration management, we will ensure that the network part of the OCCI open standard can be used in INDIGO supported CMFs: OpenStack and OpenNebula.

\section{Interfacing with the user}
\label{sec:portals}

The INDIGO architecture needs to address the challenge of guaranteeing a simple and effective final usage, both for software developers and application running.

A key component with a big impact on the end-user experience is the authentication and authorization meachanism employed to access the e-infrastructures. 

On the next layer, the possibility of using user friendly end-points in the form of graphical interfaces, that user communities may tailor to their needs is a also big plus to enhance the end-user experience.

\begin{figure}
  \centering
  \includegraphics[width=12cm]{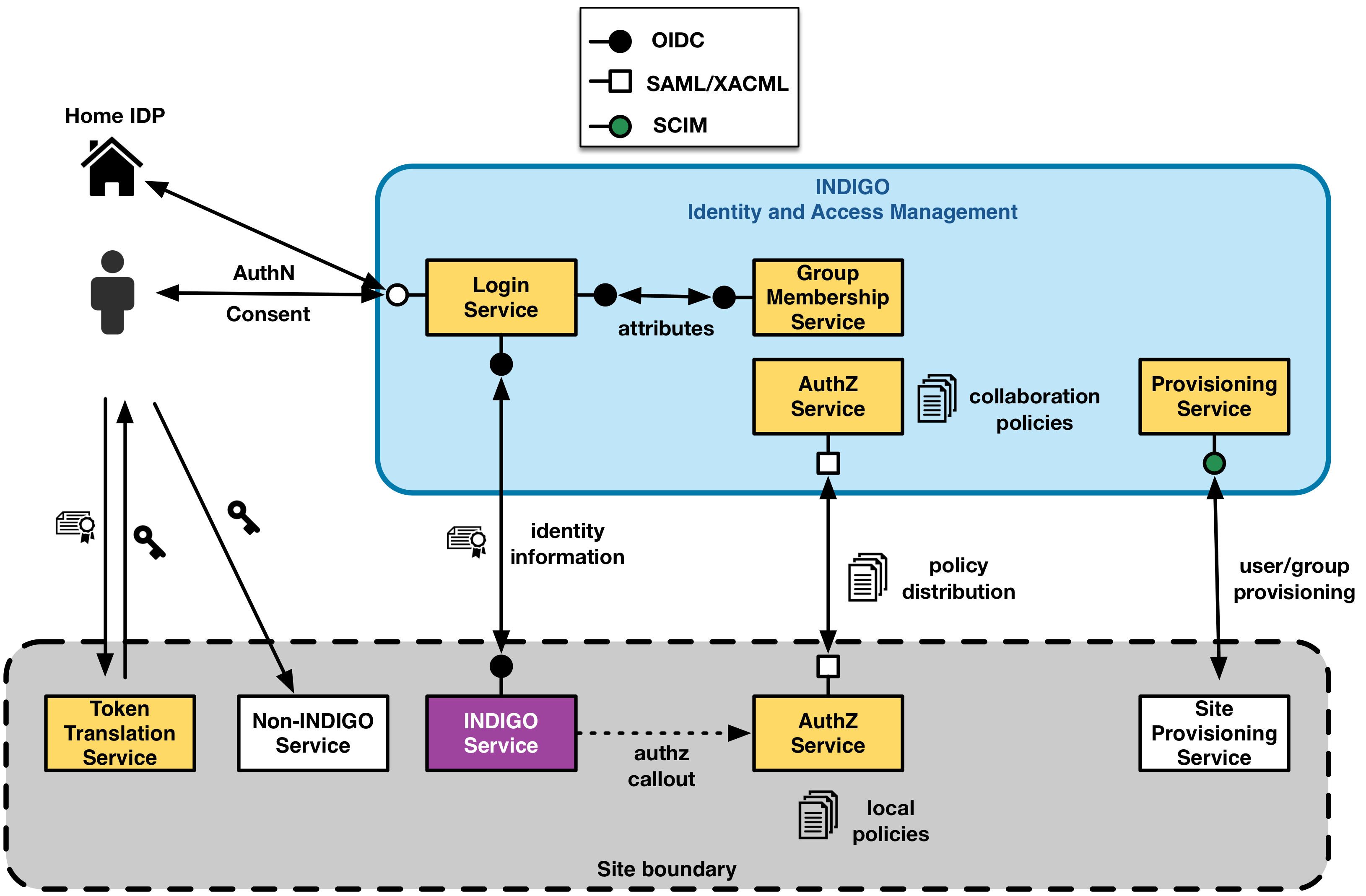}
  \caption{Components involved in the Authentication and Authorization process from the user to the infrastructure perspective}
  \label{fig:15}
\end{figure}

\subsection{Identity Management in the INDIGO Architecture}

We have designed a service providing user identity so that consistent authorization decisions can be enforced across distributed services. We can see an schema of the problem we intend to tackle in Figure \ref{fig:15}.

Today users have different digital identities (e.g., institutional credentials,
social logins, certificates) and want the ability to leverage these identities
for their scientific computing needs.

So one problem that must be solved is how to map different identities to the
same individual so that consistent authorization and accounting can be
performed at various levels of the infrastructure.

This identity harmonisation problem, which we describe in more detail in our
AAI architecture document \cite{INDIGO-AAI-ARCH}, has many aspects that need to
be tackled:

\begin{itemize}

  \item ability to authenticate users coming with different credentials

  \item ability to recognize which credentials are linked to which individuals, and provide
    a unique identifier linked to the individual (orthogonal to the different credentials used)

  \item ability to link attributes to the identity that can be used to define and enforce authorisation policies
  \item ability to provision identity information and authorisation policies to relying services

\end{itemize}

To address these points we have developed a service called {\sl Identity Access
Management} (IAM), which provides a central solution for identity
harmonisation, user authentication and authorisation.

\begin{figure}
  \centering
  \includegraphics[width=8cm]{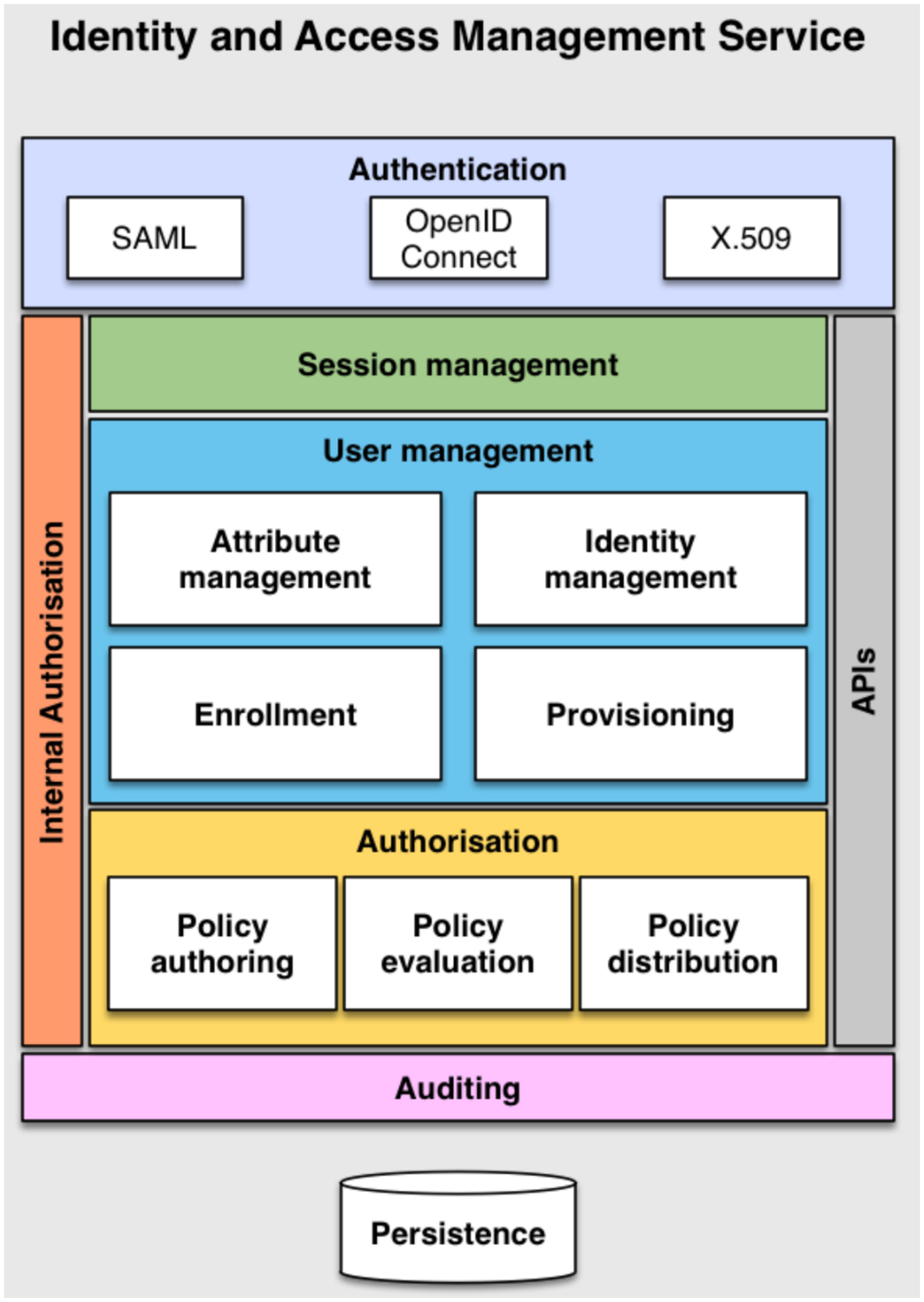}
  \caption{Architecture of the Identity Access Management INDIGO service.}
  \label{fig:16}
\end{figure}

In particular, it provides a layer where identities, enrollment, group
membership and other attributes management as well as authorization policies on
distributed resources can be managed in a homogeneous way leveraging the supported
federated authentication mechanisms (see Figure \ref{fig:16}).

The IAM service supports standard authentication mechanisms as SAML,
OpenID-connect (OIDC) and X.509. The user identity information collected in
this way is then exposed to relying services through OpenID-connect interfaces.
In a way, the IAM acts as a credential translator for relying services,
harmonizing the different identity of the users and exposing them using a
single standardized protocol. This approach simplifies integration at services
as it doesn't force each service to understand and support each authentication
mechanism used by users.

OpenID-connect was chosen as the identity layer in INDIGO for the following
reasons:

\begin{itemize}

  \item easier integration in services. Most services today are exposed via
    RESTful APIs and OpenID-connect fits naturally to that use case and many
    products that are part of the INDIGO stack already support an OIDC
    integration (e.g., OpenStack, Kubernetes);

  \item support for a dynamic computing environment: OIDC naturally supports
    dynamic client registration so that trust can be enstabilished across
    services without human intervention (but according to well-defined
    policies);

  \item support for offline access: being based on OAuth2, OIDC naturally
    supports offline access in a way which is independent of the authentication
    mechanism used;

\end{itemize}

The IAM service will represent an identity hub for INDIGO services, and
provides standard interfaces (SCIM \cite{SCIM}) for the
provisioning/deprovisioning of user and group information at relying services.

The IAM integrates with the Indigo Token Translation Service (TTS) to support
non-HTTP services (ssh, Amazon S3) and provide other needed credential
translation functionality (ie, X.509 certificate generation).

The IAM deployment model is flexible: it can be deployed centrally to
accomodate a federated infrastructure and large user communities, or locally at
a site to provide local identity harmonisation/attribute management services.

\subsection{Graphical User Interfaces}

We have developed the tools needed for the development of APIs to access the INDIGO PaaS framework. It is via such APIs that the PaaS features can be exploited via Portals, Desktop Applications or Mobile Apps. Therefore, the main goals of such endeavour from a high level perspective are to:

\begin{itemize}
\item Provide User Friendly front-ends demonstrating the usability of the PaaS services;
\item Manage the execution of complex workflows using PaaS services;
\item Develop Toolkits (libraries) that will allow the exploitation of PaaS services at the level of Scientific Gateways, desktop and mobile applications;
\item Develop an Open Source Mobile Application Toolkit that will be the base for development of Mobile Apps (applied, for example, to a use case provided by the Climate Change community).
\end{itemize}

\begin{figure}
  \centering
  \includegraphics[width=10cm]{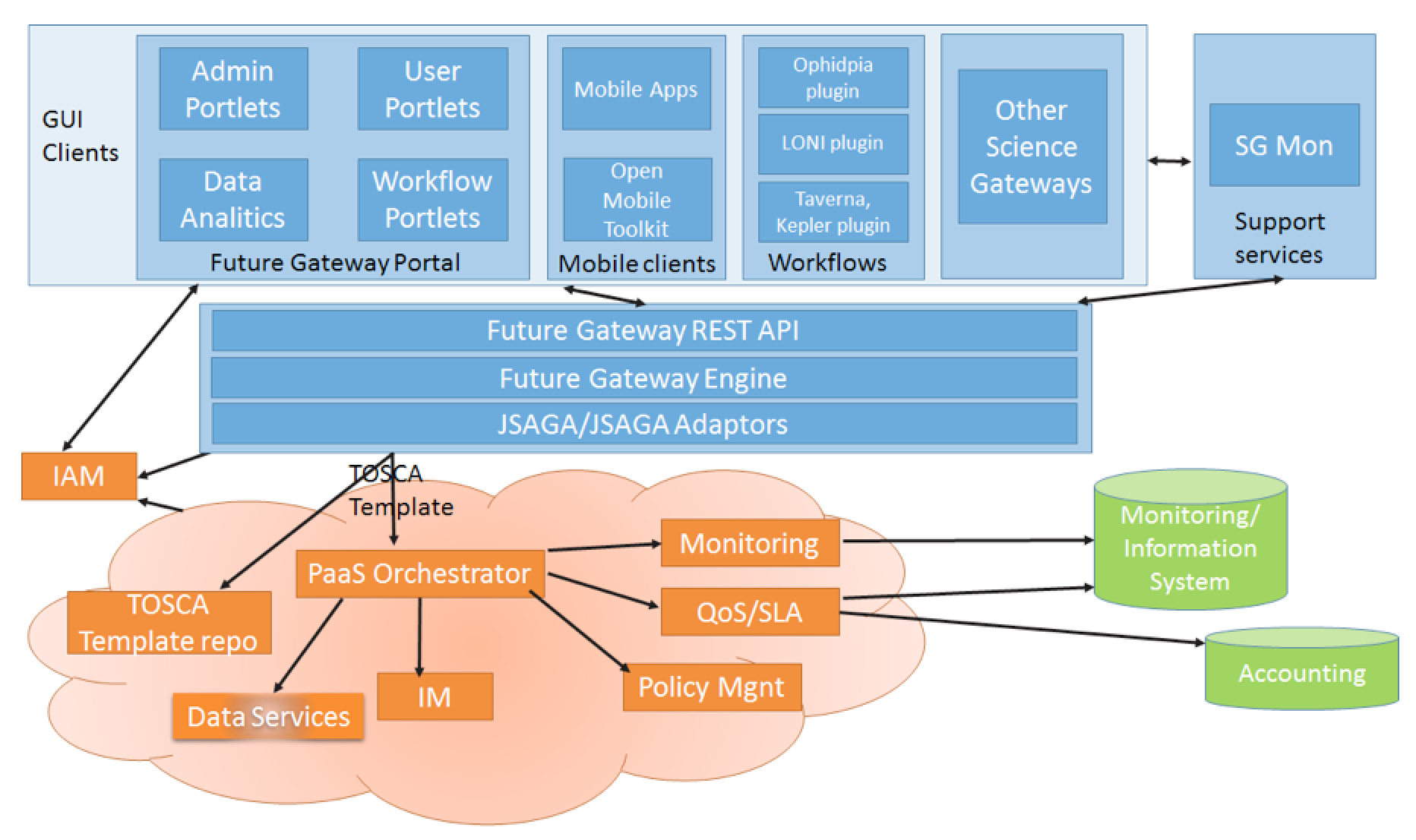}
  \caption{Interaction between components from the point of view of the user interface services}
  \label{fig:7}
\end{figure}

The architectural elements of the user interface (see Figure \ref{fig:7}) can be described as follows:

\begin{itemize}
\item FutureGateway Portal: it provides the main web front-end, enabling most of the operations on the e-infrastructure. A general-purpose instance of the Portal will be available to all users.
 
Advanced features that can be included via portlets are:

\begin{itemize}
\item Big Data Portlets for Analytics Workflows - aiming at supporting a key set of functionalities regarding big data and metadata management, as well as support for both interactive and batch analytics workflows.
\item Admin Portlet - to provide the web portal administrator with a convenient set of tools to manage access of users to resources.
\item Workflow Portlets - to show the status of the workflows being run on the infrastructure, and provide the basic operations.
\end{itemize}

\item FutureGateway Engine: it is a service intermediating the communication between e-Infrastructures and the other user services developed.  It incorporates many of the functionalities provided by the Catania Science Gateway Framework \cite{CATANIA}, extended by others specific to INDIGO. It exposes a simple RESTful API for developers building portals, mobile and desktop applications. Full details can be found in \cite{D61}.

\item Scientific Workflows Systems: these are the scientific workflow management systems orchestrating data and job flow. We have selected Ophidia, Galaxy, LONI and Kepler as the ones more demanded by the user communities. 

\item Wfms plug-ins – these are plug-ins for the Scientific Workflow Systems that will make use of the FutureGateway Engine REST API, and will provide the most common set of the functionalities. These plug-ins will be called differently depending on the Scientific Workflow system (modules, plug-ins, actors, components). 

\item Open Mobile Toolkit: these are libraries that make use of the FutureGateway Engine REST API, providing the most common set of the functionalities that can be used by multiple domain-specific mobile applications running on different platforms. Foreseen libraries include support for iOS and Android and, if required, for WindowsPhone implementations.

\item INDIGO Token Translation Service Client - The Token Translation Service client enables clients that do not support the INDIGO-token to use the INDIGO AAI architecture. 
\end{itemize}

\section{Unified Data Access}
\label{sec:data}

The main goal in providing unified data access at the PaaS level is providing users with seamless access to data. The challenge resides in hiding the complexities and heterogeneities between the infrastructures where the data is actually being stored. This is particularly more important in the case of federated data infrastructures such as the deployment of LHC data over the WLCG \cite{WLCG}.

As a matter of fact data  access  interoperability  is  currently  the  main  challenge  when  it  comes  to  federated data management. Various  Grid  and  Cloud  infrastructures  use  different  data  management  and  access technologies, either open source or proprietary. Although some solutions exist, such as CDMI, none is widely accepted and thus we need to define a unified API for users to access data from heterogeneous infrastructures in a coherent manner.

In the case of LHC data analysis in Run-2 we expect a much more heterogenous infrastructure in place, besides the WLCG Grid, resources based on cloud-like provision is also being explored. Therefore it becomes critical to provide a certain level of integration. 

\begin{figure}
  \centering
  \includegraphics[width=10cm]{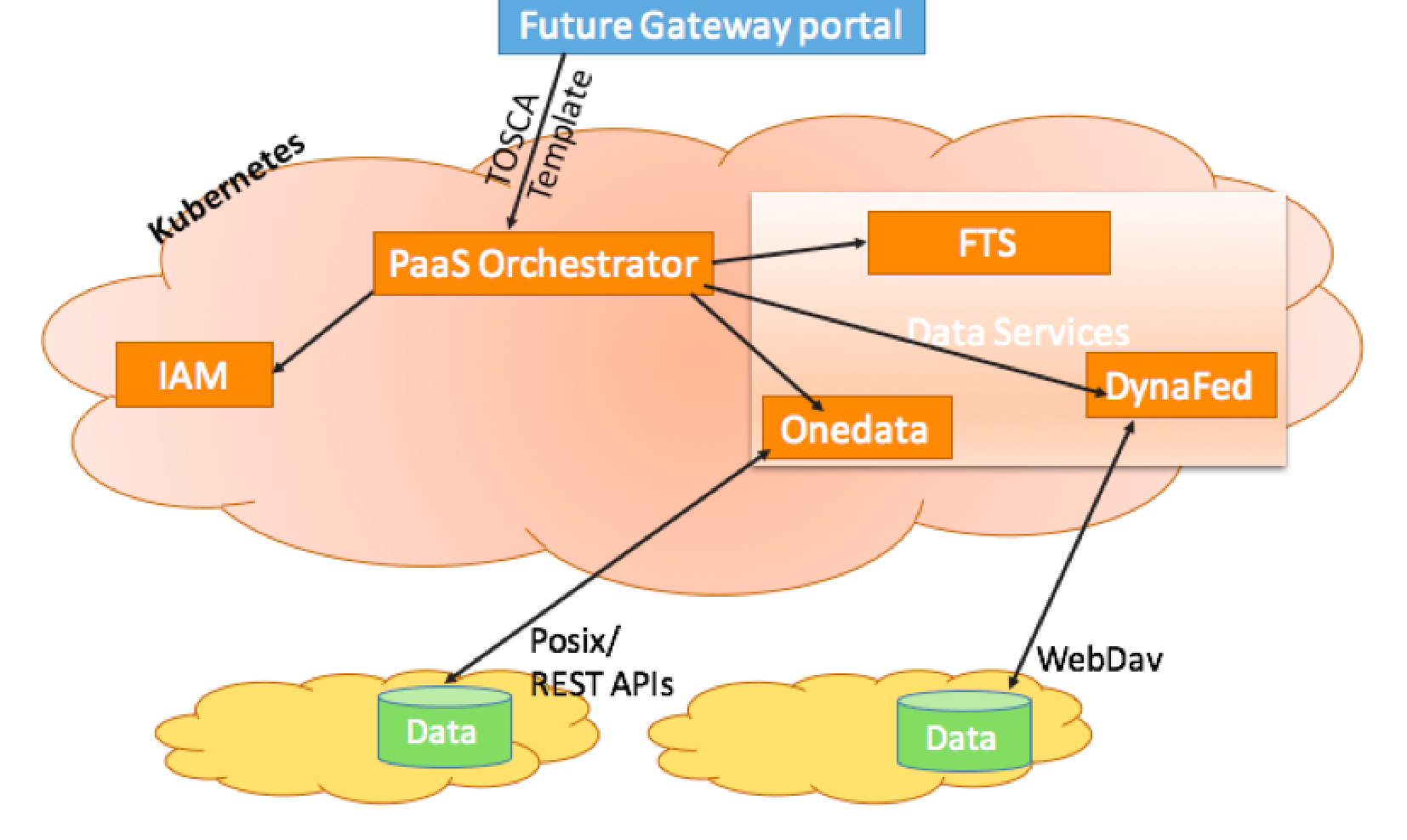}
  \caption{Data access high-level description and involved services}
  \label{fig:12}
\end{figure}

The INDIGO PaaS provides three data management services that allow accessing federated data in an unified way. Depending on how data are stored/accessible, they will be made available through a different services in a way which is transparent to the user (see Figure \ref{fig:12}). In order to access and manage data, we will exploit the interfaces provided by the infrastructure layer: 

\begin{itemize}
\item Posix and WebDAV for data access.
\item GridFTP for data transfer.
\item CDMI for the Metadata Management.
\item REST APIs to expose QoS features of the underline storage.
\end{itemize}

\subsection{OneData Service}

OneData \cite{ONEDATA} is a server-client type of service whose main purpose is to provide a unified 
federated and optimized data access based on various APIs as well as legacy POSIX protocol. It allows
transparent  access  to  storage  resources  from  multiple  data  centers simultaneously. OneData
automatically detects whether data is available on local storage and can be accessed directly, 
or whether  it  has to be fetched from remote sites.
  
Support  for  federation  is  achieved  by  the  possibility  of establishing     a     distributed     provider     registry,     where     various infrastructures  can  setup  their  own  provider  registry  and  build  trust relationship  between  these  instances,  allowing  users  from  various platforms to share their data transparently

The main dependence of OneData is on the storage providers. For OneData to work the storage providers needs to expose CDMI or S3 interfaces, and supporting POSIX access to site storage.

The architecture comprises two major components: oneprovider and oneclient. The former is installed in datacenters to provide a unified view of the site filesystems. 

The client side  connects to  the  providers which  the  user  registered  in OneData  portal, and his spaces are automatically provisioned  from these  providers.  

If data are stored on a POSIX-compliant filesystem and the site admin is willing to install the OneData gateway, data can be accessed with a more powerful graphical interface, including: 

\begin{itemize}
\item ACLs and QoS management.
\item Posix access (also remotely).
\item Web or WebDAV access.
\item Simple metadata management (based on the concept of key-value pair).
\item Data movement, replication and caching across the site of the federation.
\item Transparent local caching of data.
\item Transparent translation of object storage resources (for example available through an S3 \cite{S3} interface) to a Posix filesystem.
\end{itemize}

\subsection{FTS Service}

The FTS service will be exploited for the capability of managing third-party transfers among griftp servers. It will be used in order to import data from external gridftp servers, globus-online compatible services, etc. Support to FTS is needed in order to provide access to most WLCG storage elements. We describe here the main functionalities for completeness.

For  authentication FTS uses VOMS proxies. It optimizes transfers by monitoring the performance currently active transfers and adjusting the number of concurrent transfers on the fly.

The  users  typically  only  provide  file  origin and destination storage elements. After submitting origin and destination to the client the user can  monitor  the  progress,  query  the  status,  cancel  and  delete  transfer  requests  and  resubmit  previously  unsuccessful  requests.  
A  successful transfer from source to destination using the specified protocol 
(srm/gsiftp/http/root) can be verified by a checksum that was added during request creation.

\subsection{Dynafed Service}

Dynafed \cite{DYNAFED} is a federated namespace of distributed storage namespaces. It provides a very fast loose coupling of storage endpoints as a single name-space exposed via HTTP and WebDAV. 

This allows to have federation of  existing  storage endpoints without the need of maintaining a file catalog for global to local file name  translation.

The purpose of providing this service in the PaaS is to cover the case in which data are available only via a WebDAV gateway, they can be aggregated using DynaFed. Users will be able to read them via a federation layer regardless of where the data are really stored.

To wrap things up, in the INDIGO PaaS framework, the user will provide information about the data needed to execute the desired service/application, and how those data are to be accessed, at the level of the TOSCA Template.

Given the data requirements described in the template, the orchestrator will be able to understand if it has to request either FTS or OneData to schedule a data import/movement, or if instead it is better to move the application “close” to the data. 

At the end of this cycle the data will be available to the end user service exploiting OneData or DynaFed. This will allow also legacy application/services to be supported with their native data access approach (WebDAV or Posix). 

In summary the end user has the possibility to handle data in several ways: 

\begin{itemize}
\item Asking for an import action using a TOSCA Template exploiting FTS and Gridftp.
\item Uploading files or directories using a web interface.
\item Importing data from his desktop via a Dropbox-like tool.
\end{itemize}

\section{Conclusions and future work}
\label{sec:conclusions}

The INDIGO Architecture, based on a three level approach (User Interface, PaaS layer and Infrastructure Layer), is able to fulfil the requirements described in the introduction. 
 
 At the Infrastructure site level the architecture provides new scheduling algorithms for open source Cloud frameworks. Also very importantly, it provides dynamic partitioning of batch versus cloud resources at the site level. By implementing the Cloud bursting tools available in the architecture, the site can also have access to external infrastructures.

The infrastructure site is also enhanced with full support to containers, where the local containers repositories can as well be securely synchronized with dockerhub external repositories, facilitating enormously the automatic instantiation of applications.

From the point of view of data, the architecture is able to integrate local and remote Posix access for all types of resources: bare metal, virtual machines or containers. In particular it provides transparent mapping of object storage to Posix, and a transparent Gateway to existing filesystem (like GPFS or LUSTRE).Data access is also enhanced with the support to WebDAV, GridFTP and CDMI access.  

Using pre-defined available TOSCA templates we are also able to provide a high level of automatism for a number of scenarios. Furthermore, the PaaS scenario has the advantage that the management of the application/services has advanced capabilities related with service resilience that make it very attractive for users. For example, in the case of failure of one the services or of an application, the platform itself will take care of restarting the service or re-executing the application.

Throughout this paper we have summarized the work performed at the infrastructure, Platform and user interface level, which also includes the description and implementation of some typical use case scenarios, to provide more clarity as to what we want to support with this architectural construction.

The plan is to deliver a first release of the platform by July 2016, implementing the most important features to let users deploy their services and applications across a number of testbed provided by the INDIGO partners, and provide developers with an initial feedback.
 
In the second INDIGO release, scheduled by March 2017, we plan to also integrate advanced support for features such as moving applications to Cloud infrastructures, addressing Cloud bursting, enhancing data services and providing additional sample templates to support use cases as they are presented by scientific communities. The release cycle, roadmap and key procedures of the INDIGO software are described here \cite{ROADMAP}.

\end{document}